# First-principles based analysis of thermal transport in metallic nanostructures: size effect and Wiedemann-Franz law

Yue Hu[1], Shouhang Li[1], Hua Bao[1,*]

[1]University of Michigan–Shanghai Jiao Tong University Joint Institute, Shanghai Jiao Tong University, Shanghai 200240, P. R. China

**Abstract**

Metallic nanostructures (the nanofilms and nanowires) are widely used in electronic devices, and their thermal transport properties are crucial for heat dissipation. However, there are still gaps in understanding thermal transport in metallic nanostructures, especially regarding the size effect and validity of the Wiedemann-Franz law. In this work, we perform mode-by-mode first-principles calculations combining the Boltzmann transport equation to understand thermal transport in metallic nanostructures. We take the gold (Au) and tungsten (W) nanostructures as prototypes. It is found that when the size of nanostructures is on the order of several tens of nanometers, the electronic/phonon thermal conductivity is smaller than the bulk value and decreases with size. The phonon contribution increases in nanostructures for those metals with small bulk phonon thermal conductivity (like Au), while the phonon contribution may increase or be suppressed in nanostructures for those metals with large bulk phonon thermal conductivity (like W). By assuming that the grain boundary does not induce inelastic electron-phonon scattering, the Wiedemann-Franz law works well in both Au and W nanostructures if the Lorentz ratio is estimated using electronic thermal conductivity. The Wiedemann-Franz law also works well in Au nanostructures when the Lorentz ratio is estimated by total thermal conductivity.

[*]Author to whom correspondence should be addressed. *E-mail address*: hua.bao@sjtu.edu.cn (H. Bao).

I. Introduction

Due to the miniaturization of electronic devices, metallic nanostructures (nanofilms and nanowires) play an imperative role in electronic applications ranging from interconnects to sensors [1–3]. Since effective heat dissipation is crucial to ensure the reliability of electronic devices [4], the thermal conductivity of metallic nanostructures is crucial [5–11]. Over the past two decades, there have been many discussions on thermal transport in metallic nanostructures [9,10,12–18]. Among these discussions, there are two key issues: the size effect and the validity of the Wiedemann-Franz law.

Both experimental and theoretical evidence shows a size effect [19,20]: the thermal conductivities in the metallic nanostructures are smaller than their bulk counterparts and depend on the size of the nanostructures. This size effect is attributed to the surface scattering and grain boundary scattering of the electron/phonon [21–23]. The size effect of the electronic component has been described by Fuchs-Sondheimer theory and Mayadas-Shatzkes theory [24,25]; however, the success of these theories relies on the validity of the free electron model [7]. For the phonon component, the main strategy used in previous work is roughly estimating the phonon thermal conductivity using molecular dynamics methods or the Debye model [26–28]. However, the magnitude of phonon thermal conductivity is still under debate. For example, in gold nanostructures, some studies believe that the phonon component can be ignored [5,9,29], while other studies estimate a large phonon component, which is even larger than the electronic component [7,15]. Thus, an established understanding of the size effect on phonon thermal conductivity is still lacking.

The Wiedemann-Franz law states that the ratio of the electronic contribution of the thermal conductivity to the electrical conductivity of a metal is proportional to the temperature. The proportionality constant, known as the Lorentz ratio, is assumed to be the Sommerfeld value, $L_0 = 2.44 \times 10^{-8} W\Omega/K^2$ [30]. When the phonon contribution is negligible, the Wiedemann-Franz law provides an estimation of thermal conductivity according to the measured electrical conductivity, which is easier to be measured [30]. Meanwhile, the Wiedemann-Franz law can be used to separate the electron and phonon contributions when the total thermal conductivity is known [8]. However, the validity of the Wiedemann-Franz law in metallic nanostructures is still controversial. Several researchers reported the Lorentz ratio is several times larger than the Sommerfeld value [7,8,15,16,26], while other researchers reported the Lorentz ratio is around or smaller than the Sommerfeld value [9,10,18,31]. Moreover, it is still under debate whether the Lorenz ratio is size and temperature dependent or not [5,9,32]. Stojanovic *et al*. proposed that the deviation of the Lorentz ratio from the Sommerfeld value is related to the neglect of phonon contribution and the inelastic electron-phonon scattering [28]. A quantitative understanding of phonon contribution and inelastic electron-phonon scattering is desired to understand the Wiedemann-Franz law in metallic nanostructures.

In this work, we tackle these two issues using a theoretical calculation combining first-principles calculations and Boltzmann transport equation. We take gold (Au) and tungsten (W) as testing materials, as their transport properties are widely studied in experiments [9,15,33,34]. Furthermore, the two metals have typical characteristics from a theoretical perspective (Au has a near spherical Fermi surface and W has a complicated Fermi surface [35,36]). Different from previous theoretical calculations [24,25,28], we consider the intrinsic electron and phonon properties using rigorous mode-by-mode first-principles calculations [36–38], instead of applying assumptions such as the free electron assumption and Debye model for the phonon [28]. Combining the Boltzmann transport equation with the intrinsic electron and phonon properties, the thermal transport properties (thermal conductivity and Lorentz ratio) of metallic nanostructures can be predicted.

## II. Method and Simulation details

In this section, we provide our calculation framework: first-principles calculations combining the Boltzmann transport equation for predicting the thermal conductivity of metallic nanostructures. Since our framework is based on the Boltzmann transport equation, we restricted our focus to length scales beyond ten nanometers, for which the quantum confinement effects of the electron or phonon can be ignored [39,40]. Both the electron and phonon are involved in the thermal transport of metallic nanostructures. The electron and phonon in nanostructures not only encounter intrinsic scattering processes in ideal bulk [35,37], but also encounter external scattering processes including surface scattering and grain boundary scattering [24,25].

a)  **Surface scattering and grain boundary scattering**

Both surface scattering and grain boundary scattering are important in the phonon and electron transport in metallic nanostructures [24,25]. Schematics of these scattering processes as well as the typical structure of a nanofilm/nanowire with thickness/diameter $H$ and grain size $D$ are shown in Fig.1. The grain boundary is assumed to be perpendicular to the surface, as observed in the experimental samples [22,25]. The proportionality between the grain size $D$ and the thickness $H$ is usually observed in the experiments, and the proportionality constant is usually in the range of 2-0.5 [22,25,41].

To describe the surface scattering, we adopt the idea of Fuchs' model: some of the carriers specularly reflected by the surface, while others diffusely reflected by the surface (Fig. 1). The percentage of specularly reflected carriers is specularity $p$. This model is also known as the partial diffuse-partial specular boundary condition, which is widely used in surface scattering modeling [42–44]. The specularity $p$ can be obtained by Soffer's model [45,46],

$$p = \exp\left(-16\pi^2\eta^2 / \Lambda^2 \cos^2\theta\right),$$

where $\eta$ is the roughness of the surface, and $\Lambda$ is the wavelength of the electron or phonon. $\theta$ is the incident angle. The roughness $\eta$ of the surface can be measured, and the reported values are in the order of 1 nm [22,47,48]. According to Soffer's model, when the roughness is on the order of 1 nm, the specularity is nearly zero (pure diffuse surface scattering) for all electrons and phonons in both Au and W. Thus, we assume a diffuse surface scattering for all carriers in subsequent sections.

The scattering around the grain boundary is complex and similar to the scattering at the interface [49,50]. To simplify the complex grain boundary scattering, we adopt the idea of Mayadas' model: some of the carriers transmit across the grain boundary, while others are reflected at the grain boundary. The percentage of reflected carriers is the reflection coefficient $R$ [25]. The accuracy of this model depends on the reflection coefficient. As far as we know, it is difficult to theoretically obtain the reflection coefficient $R$, which is usually obtained by fitting the experimental results [27,41,51]. The range of reported $R$ for electrons is 0.2-0.8 for polycrystalline nanostructures [8,28,29,52]. The large range may be due to the fact that $R$ is related to the configuration of the grains, which varies for different samples. The frequency of the phonon is much smaller than the frequency of the electron. Therefore, the reflection coefficients of phonon modes can be assumed to be zero, i.e., ignoring the grain boundary scattering of phonons [50,53,54].

Since the structural scatterings are usually treated as elastic scattering [55–57], we assume the charge and heat transport of the electron have the same specularity and reflection coefficient.

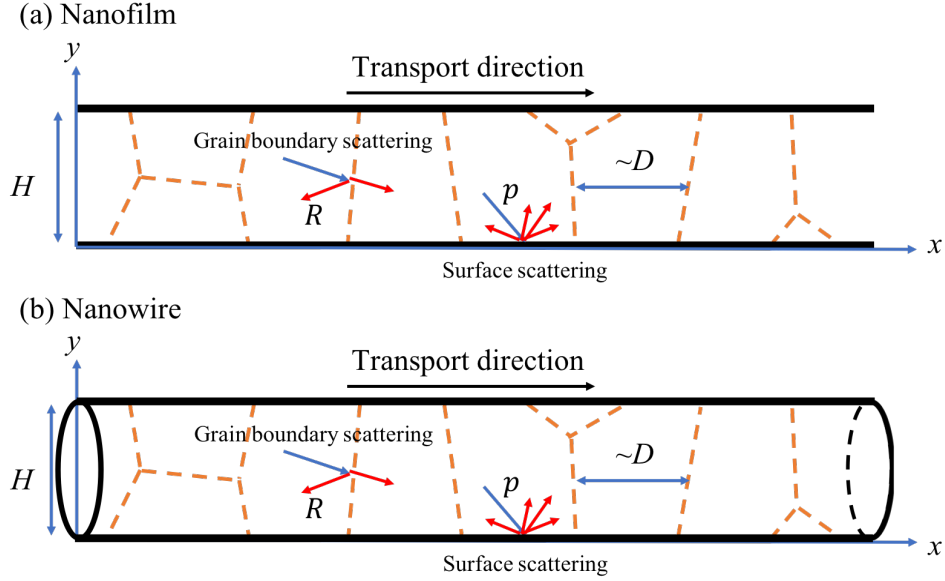

Fig.1. Surface scattering and grain boundary scattering in a typical (a) nanofilm and (b) nanowire. The thickness of the nanofilm is $H$ and the average grain size is $D$. The diameter of the nanowire is $H$ and the average grain size is $D$. The specularity of carriers at the surface is $p$. The reflection coefficient of carriers at the grain boundary is $R$.

b) **Thermal conductivity: Boltzmann transport equation model**

By solving the Boltzmann transport equation with the relaxation time approximation [19], the phonon and electronic thermal conductivity in the transport direction of a metallic nanostructure (nanofilm or nanowire) can be calculated as

$$\kappa = \sum_\lambda c_\lambda \left\| v_{x,\lambda} \right\|^2 \tau_\lambda S_\lambda, \qquad (1)$$

where $\kappa$ is the electronic thermal conductivity $\kappa_e$ or phonon thermal conductivity $\kappa_{ph}$. For the phonon thermal conductivity, $\lambda = (\mathbf{q},\upsilon)$ denotes the phonon mode with polarization $\upsilon$ and wave vector $\mathbf{q}$. For the electronic thermal conductivity, the $\lambda = (\mathbf{k},i)$ denotes the electron with wave vector $\mathbf{k}$ and band index $i$. $c_\lambda$ is the volumetric specific heat of the phonon or the electron. $v_{x,\lambda}$

denotes the component of the phonon group velocity or the electron velocity in the transport direction. The $\tau_\lambda$ denotes the intrinsic phonon relaxation time or the intrinsic electronic thermal transport relaxation time. The $S_\lambda$ denotes the "suppression function" for phonons and electrons, which captures the size dependence of the thermal conductivity. For bulk materials, the $S_\lambda$ equals to 1, and Eq. (1) is reduced to the expression for bulk materials [58]. Thus, Eq. (1) can be simplified as $\kappa = \sum_\lambda \kappa_{bulk,\lambda} S_\lambda$. The $S_\lambda$ for nanofilms and nanowires are related to surface scattering and grain boundary scattering. For the nanofilm, the $S_\lambda$ is expressed as

$$S_\lambda(L,D,\tau_\lambda,v_\lambda,p_\lambda,R) = \frac{\tau'_\lambda}{\tau_\lambda} \frac{1}{H} \int_0^H \left(1 - \frac{1-p_\lambda}{1-p_\lambda \exp(-L/\tau'_\lambda |v_{y,\lambda}|)} \exp\left(-\frac{|y-y_B|}{\tau'_\lambda |v_{y,\lambda}|}\right)\right) dy, \quad (2)$$

where the $\tau'_\lambda$ is the relaxation time considering the intrinsic scattering processes (electron-phonon, phonon-phonon scattering) and grain boundary scattering process, and is expressed as

$$\frac{1}{\tau'_\lambda} = \frac{1}{\tau_\lambda} + \frac{\alpha}{2\tau_\lambda} \frac{v_\lambda}{|v_{\lambda,x}|}, \alpha = \frac{v_\lambda \tau_\lambda}{D} \frac{R}{1-R}.$$

When there is no grain boundary scattering, the $\tau'_\lambda$ is equal to the intrinsic relaxation time $\tau_\lambda$. The electron/phonon properties including velocity, heat capacity, and intrinsic relaxation time are obtained by first-principles calculations. The detailed derivation of Eq. (2) as well as the equation for nanowires are presented in the Supplementary Material [59].

The formula above can also be used in electrical conductivity calculation, i.e., $\sigma = \sum_\lambda \sigma_{bulk,\lambda} S_\lambda$, which will not be discussed here.

c)  **Intrinsic phonon and electron properties: first-principles calculations**

For phonons, the volumetric specific heat can be obtained by $c_{v,\lambda} = \frac{\hbar \omega_\lambda}{V} \frac{\partial n^0_\lambda}{\partial T}$, with $n^0_\lambda$ the Bose-Einstein distribution function, $V$ the volume of the primitive cell, $\omega_\lambda$ the phonon frequency, and $T$ the temperature. The phonon group velocity can be obtained by $\mathbf{v}_\lambda = \frac{\partial \omega_\lambda}{\partial \mathbf{q}}$. The phonon relaxation time can be obtained using the Matthiessen's rule [60] considering both phonon-phonon and phonon-electron scattering [61], namely $1/\tau_\lambda = 1/\tau_{pp} + 1/\tau_{pe}$, where

$1/\tau_{pp}$ is the scattering rate with respect to three-phonon interactions [60], and $1/\tau_{pe}$ is the scattering rate with respect to phonon-electron interactions [62]. For electrons, the volumetric specific heat can be obtained by $c_{v,n\mathbf{k}} = -\frac{n_s}{VT}(\varepsilon_{n\mathbf{k}} - \mu)\frac{\partial f^0_{n\mathbf{k}}}{\partial \varepsilon_{n\mathbf{k}}}$, with $f^0_{n\mathbf{k}}$ the Fermi-Dirac distribution function, $V$ is the volume of the primitive cell, $n_s$ denotes spin degeneracy, $\varepsilon_{n\mathbf{k}}$ is the electron energy, and $\mu$ is the chemical potential. The electron group velocity can be obtained by $\mathbf{v}_{n\mathbf{k}} = \frac{1}{\hbar}\frac{\partial \varepsilon_{n\mathbf{k}}}{\partial \mathbf{k}}$. For pure metals at intermediate and high temperatures, the electron relaxation time is limited by electron-phonon scattering and can be obtained using Fermi's golden rule [63]. In particular, we rigorously consider the elastic/inelastic electron-phonon scatterings and their effects on transport coefficients by distinguishing the electron relaxation time for heat and charge transport. More computational details of the intrinsic phonon and electron properties (for both heat and charge transport) from first-principles methods have been described previously [37].

We carried out the first-principles calculations using the Quantum Espresso [64] package to predict electron transport. The types of Local Density Approximation (LDA) [65] and Generalized Gradient Approximation (GGA) [66] exchange-correlation functional are used for Au and W, respectively. The phonon-phonon scattering rate is evaluated by ShengBTE package [67]. The phonon-electron scattering rate and electron-phonon scattering rate are evaluated by ELETRON PHONON WANNIER (EPW) package [68]. The $\mathbf{k}$-point and $\mathbf{q}$-point meshes are set to be $80\times80\times80$ and $60\times60\times60$ for Au, while $100\times100\times100$ and $40\times40\times40$ for W to sample the Brillouin zone.

### d) Bulk thermal conductivity and mean free path

We present the bulk thermal conductivity for Au and W in Fig. 2(a), calculated by Eq. (1) (the suppression function $S_\lambda$ is equal to 1). The calculated bulk thermal conductivity matches well with the experimental results [69](the error is less than 10%). In bulk Au, the phonon contribution is negligible (less than 1% of the total thermal conductivity). In contrast, in bulk W, the phonon contribution is more than 30% of the total thermal conductivity.

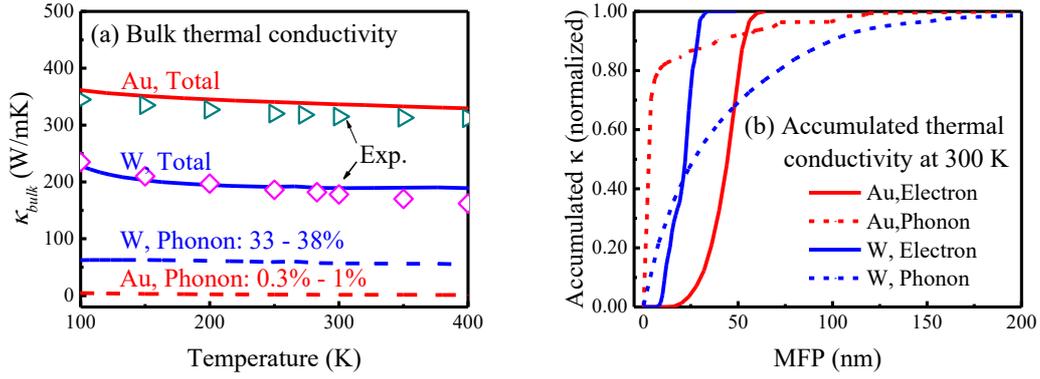

Fig. 2. (a) Bulk thermal conductivity of Au and W. The experimental data are taken from Ref. [69]. (b) The normalized thermal conductivity accumulation with respect to carrier mean free path (MFP) in Au and W at 300 K.

The thermal conductivity accumulation [70] with respect to the carrier mean free path at 300 K is shown in Fig. 2(b). This property is crucial for our mode-by-mode calculation. This also shows that our model is more accurate than previous studies, which usually adopt the averaged electron MFP from the free electron model and phonon thermal conductivity from molecular dynamics or the Debye model [15,28]. In general, the range of electron mean free path is much narrower than that of phonon mean free path for both Au and W. In Au, 80% of the phonon thermal conductivity is contributed by phonons with mean free path less than 10 nm, which is smaller than the mean free path of the electron. In W, above 50% of the phonon thermal conductivity is contributed by phonon with mean free path less than 25 nm, which is larger than the mean free path of electron. At lower temperatures, the mean free path of carriers becomes larger [35,38], which is not shown here for simplicity.

### III. Size effect

In this section, we study the size effect of the total thermal conductivity, electronic thermal conductivity, phonon thermal conductivity, and the phonon contribution. Although we only present the results for nanofilms in the main text, the results for nanowires are presented in the Supplementary Material [59]. All the conclusions about nanofilms can also be applied to nanowires.

a) **Total thermal conductivity**

In this subsection, we study the size effect of total thermal conductivity in Au and W nanofilms (including single-crystalline and polycrystalline nanofilms) at 300 K and 100 K. Previous studies demonstrated that the electrical conductivity in W single-crystalline nanofilms is related to the transport orientation [59,60]. Thus, we study the single-crystalline for [100] and [110] transport orientations. Polycrystalline nanostructures always relate to multiple orientations and thus we assume a random distribution of orientations in calculating grain boundary scattering (in polycrystalline nanofilms). As described in Sec. II(a), we assume the pure diffuse surface scattering. For polycrystalline nanofilms, we present the data for grain size $D$ equal to the thickness $H$ and electron reflection coefficient $R$ equal to 0.5. Other $D/H$ values and electronic $R$ values will be discussed later.

The normalized total thermal conductivity (over the bulk total thermal conductivity) values are shown in Fig. 3. For all nanofilms, the thermal conductivity is smaller than 90% of the bulk value for sizes ranging from 400 nm to 20 nm. Since the mean free path of carriers is on the order of 10 nm, the size effect is obvious when the size is less than 10 times the mean free path [24,25,71]. When the size decreases, the thermal conductivity of the nanofilm decreases. At 300 K, the thermal conductivity reduces to 20% of the bulk value for the 20 nm Au polycrystalline nanofilm and 35% of the bulk value for the 20 nm W polycrystalline nanofilm (Fig. 3(a) and (c)). Since the mean free path of carriers increases with temperature decreasing, the reduction of thermal conductivity is larger at 100 K and reaches 90% for the 20 nm Au polycrystalline nanofilm and 80% for the 20 nm W polycrystalline nanofilm. Compared with polycrystalline nanofilms, the reduction is smaller in the single-crystalline nanofilm (50% reduction for the 20 nm Au nanofilm and 40% reduction for the 20 nm W nanofilm at 300K) due to the absence of electron grain boundary scattering (Fig. 3(b) and (d)). In single-crystalline nanofilms, the thermal conductivity is transport orientation dependent. The difference between the two orientations increases when the size decreases. However, compared with the large reduction of thermal conductivity from bulk values (around 50% at 20 nm), the difference between these two orientations is not large (less than 10% of the bulk thermal conductivity at 20 nm). This anisotropy of thermal conductivity is similar to the anisotropy of electrical conductivity observed in previous work, which is due to the anisotropy of the Fermi surface [72].

In the following subsections, we analyze the size effect of the electron component and the phonon component separately. Since the polycrystalline nanofilms are usually fabricated in experiments and contain both surface and grain boundary scattering [8], we will only present the data for polycrystalline nanofilms in the remaining parts.

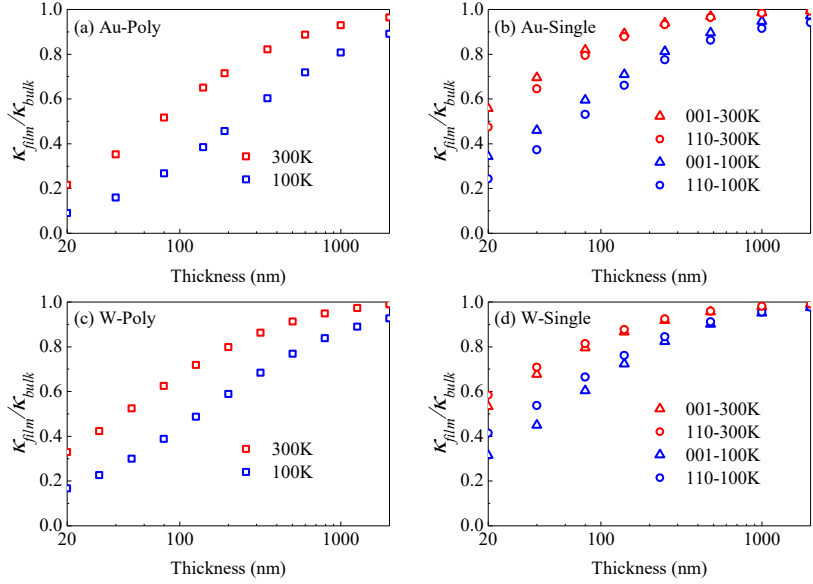

Fig. 3. Thickness $H$ dependence of normalized thermal conductivity over bulk thermal conductivity for the (a) polycrystalline Au nanofilm, (b) single-crystalline Au nanofilm, (c) polycrystalline W nanofilm, and (d) single-crystalline W nanofilm.

b) **Electronic thermal conductivity**

As described in Sec. II(a), we assume pure diffuse surface scattering. The ratio of the grain size to the thickness $D/H$ ranges from 0.5 to 2. The electron reflection coefficient $R$ ranges from 0.2 to 0.8. The range of corresponding thermal conductivity at 300 K is shown in Fig. 4 as the shadow region. We also present the typical values corresponding to $D/H=1$ and $R=0.5$ in Fig. 4 as dots. At 300 K, the electronic thermal conductivity is reduced to 90% of the bulk value when the size is smaller than 400 nm for Au and 200 nm for W, because the electron mean free path is around 40 nm for Au and 20 nm for W. When the thickness decreases, the electronic thermal conductivity decreases. At 300K, the thermal conductivity reduces to 20% for the 20 nm Au nanofilm and 40% for the 20 nm W nanofilm. Because the mean free path of electrons increases with temperature decreases, this reduction is larger at 100 K and around 90% for Au and W at 20 nm. When $D/H$ decreases or *R* increases, the electronic thermal conductivity decreases due to the larger grain boundary scattering. At 300 K, for *D/H* ranges from 0.5 to 2 and *R* ranges from 0.2 to 0.8, when the thickness is 20 nm, the reduction of electronic thermal conductivity ranges from 55% to 98% for Au nanofilm and ranges from 40% to 90% for W nanofilm.

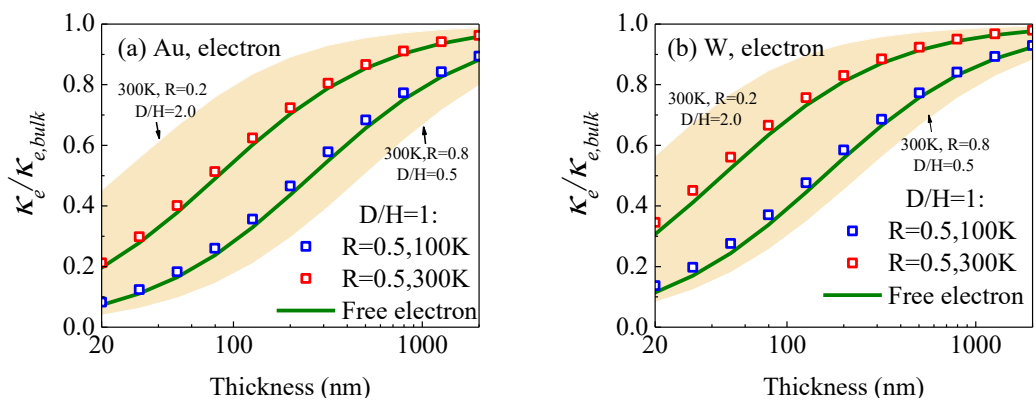

Fig. 4. Thickness *H* dependent electronic thermal conductivity over bulk electronic thermal conductivity for the (a) Au nanofilm and (b) W nanofilm.

The free electron model has been widely used in previous theoretical works [24,25]. However, whether or not it is good at metallic nanostructure properties prediction remains unknown. In this model, one assumes that the velocities of all electrons are the Fermi velocity ($1.39 \times 10^6$ m/s for Au and $6.1 \times 10^5$ m/s for W [73,74]); the Fermi surface is spherical; and the mean free path of all electrons is the same, which is the averaged mean free path calculated by $l_F = \kappa_{e,bulk} / \left( \frac{1}{3} c_e v_F \right)$). The average mean free path of electrons is 45 nm for Au and 22 nm for W at 300 K (The average mean free paths from 100 K to 400 K are presented in Supplementary Material [59]). We also calculate the electronic thermal conductivity using the free electron model. The results from the free electron model match well with the mode-by-mode results (error less than 10% for both Au and W, Fig. 4). This observation is also valid for other *D/H* values, *R* values, and other temperatures between 100 K and 400 K, which are not shown here for simplicity. This phenomenon is due to the fact that the range of electron mean free paths is not large for both W and Au (Fig. 2).

c) **Phonon thermal conductivity**

In this subsection, we discuss the size effect of phonon thermal conductivity. As described in Sec. II(a), we assume pure diffuse surface scattering and ignore the grain boundary scattering of phonons. The reduction of the phonon thermal conductivity for the 20 nm Au nanofilm is only 17% at 300 K (Fig. 5(a)), since the mean free path of the phonon is only around 2 nm (Fig. 2(b)). The reduction reaches 30% at 100 K due to the increase in the mean free path. In contrast, the reduction in phonon thermal conductivity for W is much larger (Fig. 5(b)). The thermal conductivity is reduced to 90% of the bulk value when the size is smaller than 700 nm at 300 K. This phenomenon is related to the large mean free path of the phonon in W, which spans from several nanometers to 200 nm. The reduction reaches 70% for the 20 nm W nanofilm. The reduction increases further at lower temperatures.

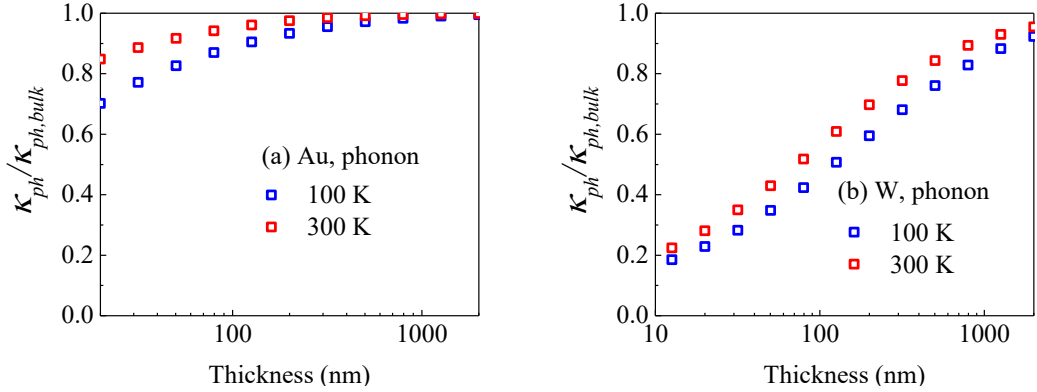

Fig. 5. Thickness $H$ dependent phonon thermal conductivity over bulk phonon thermal conductivity for the (a) Au nanofilm and (b) W nanofilm.

Apart from the value of phonon thermal conductivity, the percentage of phonon thermal conductivity contributing to the total thermal conductivity (phonon contribution) also receives attention [35,36]. Thus, we present the percentage of phonon thermal conductivity in Fig. 6. We also present the range of the phonon contribution as the shadow region and the typical values corresponding to $D/H=1$ and electron $R=0.5$ as dots. In general, the $\kappa_{ph}/\kappa_{total}$ in nanofilms can be much different from that in bulk. For Au, the phonon contribution is larger in nanofilms than in bulk and increases with the size decreases (Fig. 6(a)). For $D/H=1$ and electron $R=0.5$, the $\kappa_{ph}/\kappa_{total}$ increases to 1% for the 20 nm Au nanofilm at 300K. This percentage further increases with temperature decreases, the grain size decreases and the reflection coefficient of electrons increases. At 100 K, the $\kappa_{ph}/\kappa_{total}$ can reach up to 12%, which is much larger than 1% in bulk. The increment of the phonon contribution in Au nanofilms is due to the smaller size effect of the phonon part than that of the electron part, which is caused by two facts: (1) the phonon mean free path is smaller than the electron mean free path in Au (Fig. 2(b)); (2) the phonon reflection coefficient is smaller than the electron reflection coefficient. At lower temperatures, the larger size effect further increases the phonon contribution. In contrast, for W, the $\kappa_{ph}/\kappa_{total}$ is larger in some nanofilms and smaller in other nanofilms compared with in bulk W (Fig. 6(b)). For $D/H=1$ and electron $R=0.5$, the percentage is smaller in the 20 nm W nanofilm (30%) than in the bulk value (35%) at 300 K. However, at 100 K, the percentage is larger in 20 nm W nanofilm (45%) than in the bulk value (33%). The reason for this is the trade-off between the larger phonon mean free path versus the electron mean free path (Fig. 2(b)) and the smaller phonon reflection coefficient versus the electron reflection coefficient. The former causes a larger size effect in phonons than in electrons, while the latter causes a smaller size effect in phonons than in electrons. This trade-off also varies at different temperatures. When the grain size decreases and the reflection coefficient of electrons increases (high granular nanostructures), the percentage increases and can reach above 80% at 100 K, which is much larger than 33% in bulk W. When the grain size increases and the reflection coefficient of electrons decreases (low granular nanostructures), the percentage decreases and can be less than 30% at 300 K, which is smaller than 35% in bulk W.

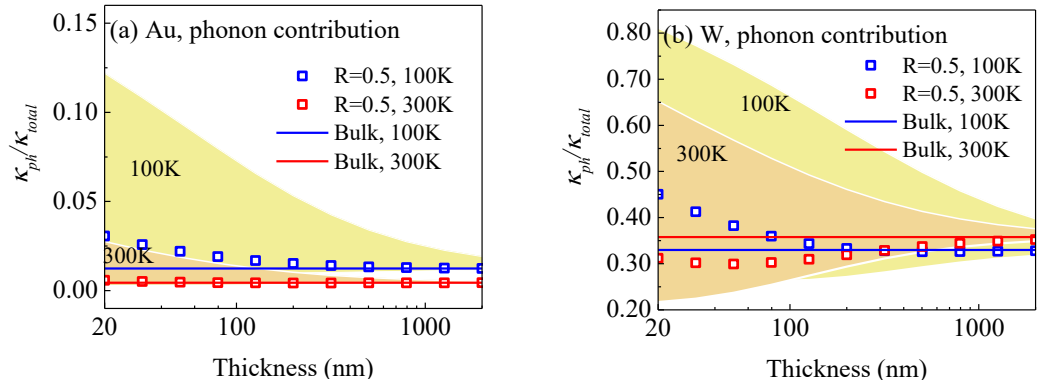

Fig. 6. Thickness $H$ dependent percentage of phonon thermal conductivity contributing to the total thermal conductivity for the (a) Au nanofilm and (b) W nanofilm.

## IV. Validity of the Wiedemann-Franz law

In this section, we study the validity of the Wiedemann-Franz law. We first discuss the Lorentz ratio estimated by the electronic thermal conductivity: $L_e = \kappa_e / \sigma T$. Then, we discuss the Lorentz ratio estimated by the total thermal conductivity: $L = \kappa_{total} / \sigma T$. Although we only present the results for nanofilms in the main text, the results for nanowires are presented in the Supplementary Material [59]. All the conclusions about nanofilms can also be applied to nanowires.

When considering the electronic thermal conductivity, the Lorentz ratios $L_e$ of bulk Au and bulk W are smaller than the Sommerfeld value (Fig. 7). When the temperature is lower, the discrepancy gets larger. This phenomenon is due to an increase in inelastic electron-phonon scattering at the intermediate temperature regime [37]. This inelastic electron-phonon scattering restrains thermal transport more than charge transport. For the nanostructures (nanofilms), due to the additional elastic scattering induced by surface or grain boundary scattering, the ratio of inelastic scattering decreases. Thus, the Lorentz ratio $L_e$ of the nanostructure is closer to the Sommerfeld value when the size effect increases, i.e., the thickness and grain size decrease. (Fig. 7). For Au, the Lorentz ratio of a nanofilm matches well with the Sommerfeld value in the whole temperature range we studied (100 K-400 K). In contrast, for W, the Lorentz ratio of a nanofilm is still smaller than the Sommerfeld value, since the inelastic scattering in W is significant.

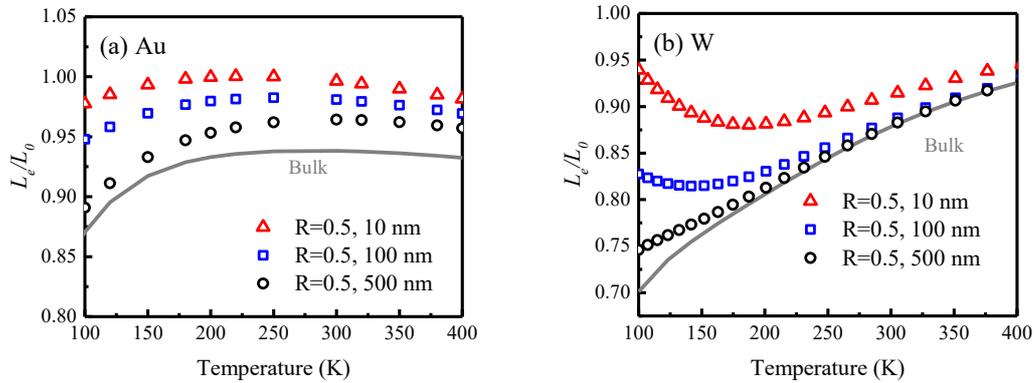

Fig. 7. Temperature dependent Lorentz ratio $L_e$ for the (a) Au nanofilm and (b) W nanofilm. The Lorentz ratio is normalized by Sommerfeld value, as $L_0 = 2.44 \times 10^{-8} \, W\Omega/K^2$

Since the electronic thermal conductivity is difficult to obtain in experiments, the Lorentz ratio is sometimes estimated by the total thermal conductivity $L = \kappa_{total}/\sigma T$. The Lorentz ratio $L$ can be larger than the Sommerfeld value due to the additional phonon contribution (Fig. 8). Since the phonon contribution in nanofilms is different from that in bulk metals, the increment of Lorentz ratio $L$ is also different from that in bulk metals. In Au, due to the increment of phonon contribution and the decreased ratio of inelastic scattering, the Lorentz ratio $L$ is larger in nanofilms than in bulk Au. For W, since the phonon contribution in nanofilms has a large range (Fig. 6), the Lorentz ratio $L$ in nanofilms also has a large range, and the upper limit can reach 2.4 times of the Sommerfeld value.

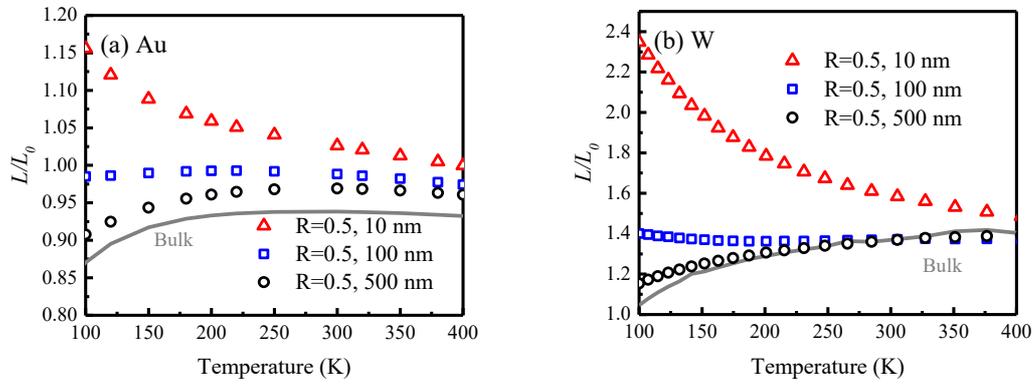

Fig. 8. Temperature dependent Lorentz ratio $L$ for the (a) Au nanofilm, (b) W nanofilm. The Lorentz ratio is normalized by the Sommerfeld value, as $L_0 = 2.44 \times 10^{-8} \, W\Omega/K^2$

The Au polycrystalline nanofilm has also been studied in experimental studies [5,7,9,27]. The reported Lorentz ratios $L$ are shown in Fig. 9. The reported Lorentz ratios $L$ spans a large range from 3 times to 0.1 times of the Sommerfeld value. In these works, the $L$ larger than the Sommerfeld value is ascribed to the phonon contribution [7,27], and the $L$ smaller than the Sommerfeld value is explained by the inelastic electron-phonon scattering [5,9]. The Bloch-Grüneisien (BG) model [57,75,76] is also applied to fit the deviation of Lorenz ratio $L_e$ in nanostructures previously [5,10,18]. However, the BG model is not a quantitative model due to the assumptions in it, such as the Debye phonon spectrum and electrons are only scattered by longitudinal acoustic phonons [37]. In our model, we rigorously consider the phonon contribution and include the intrinsic inelastic electron-phonon scattering. We also present the range of our predicted results in Fig. 9. The upper limit of the range is the Sommerfeld value plus the phonon contribution (set as 13%, since the percentages we obtained are below 13%). The lower limit is the bulk Lorentz ratio. Some experimental results [5,7,9,27] exceed the range of predicted results. Since we assume that the existence of grain boundaries does not induce further inelastic electron-phonon scattering, one possible reason for the deviation is that the high granular nature of the experimental sample induces further inelastic electron-phonon scattering [9]. The inelastic scattering can induce the Lorentz ratio value to be the Sommerfeld value multiplied by an efficiency factor, which is within the theoretical limit between 0-2 [37,55,57]. This also indicates that the mechanism of electron and phonon scattering at grain boundaries should be further studied. Finally, we note that only a single set of experimental data (Ref. [29]) observed that the Lorentz ratio value exceeds the range of 0-2 times and reaches 3 times the Sommerfeld value, which cannot be explained by inelastic scattering, and the mechanism should be further analyzed.

A recent careful experimental measurement reveals that when excluding the contact resistance, the Lorentz ratio of silver nanowires is close to the Sommerfeld value [11], which is consistent with our findings (silver also has a small phonon contribution, which is similar to Au). It indicates that there may be some uncertainties in earlier experimental measurements due to the challenge in dealing with contact resistance. Thus, to fully understand these effects, further experiments need to be carefully performed.

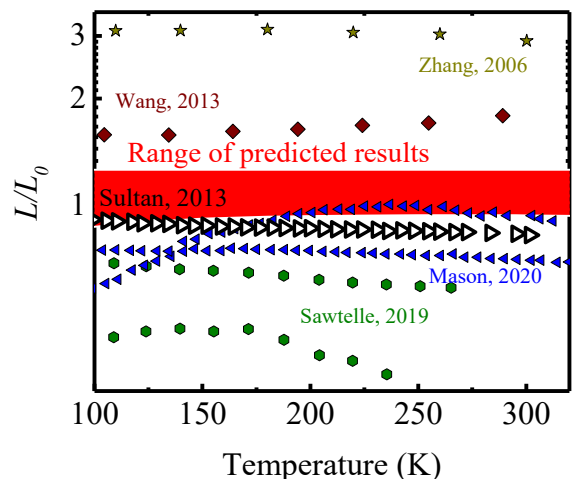

Fig. 9. Reported Lorentz ratios for Au nanofilms (normalized by the Sommerfeld value) in previous experimental studies [5,7,9,27,77].

## V.  Conclusions

In this study, we use the framework of first-principles calculations combining the Boltzmann transport equation to study the thermal transport in metallic nanostructures. Within this framework, the mode-by-mode electron and phonon properties are rigorously considered. The size effect and the validity of the Wiedemann-Franz law in Au and W nanostructures are studied.

For the size effect, we show that the thermal conductivity is smaller than the bulk value when the size is in the range of 10-1000 nm. Both the phonon and electronic thermal conductivity decrease with the size. Specifically, the size effect of the electronic thermal conductivity can be well described by the free electron model with an error less than 10%. However, the phonon contribution to the total thermal conductivity is complicated in metallic nanostructures, as it relates to the metals, temperatures, grain properties, and sizes. For Au, the phonon contribution in a thin film is larger than that in bulk metal and can reach up to 15%, which is much larger than 1% in bulk Au. For W, the phonon contribution is in a large range (20%-80%), which can be either larger or smaller than the bulk value (~30%). Roughly, the phonon contribution increases in nanostructures for those metals with small bulk phonon thermal conductivity (like Au), while it can be suppressed in nanostructures for those metals with large bulk phonon thermal conductivity (like W).

By assuming that the grain boundary does not induce inelastic electron-phonon scattering, the Lorentz ratio of nanostructures estimated by the electronic thermal conductivity is between the bulk value and the Sommerfeld value. The Lorentz ratio is closer to the Sommerfeld value when size decreases. In Au and W nanostructures, the Wiedemann-Franz law works well. The Lorentz ratio estimated by total thermal conductivity can be larger than the Sommerfeld value and can reach a large value due to the large phonon contribution. Some reported Lorentz ratios for Au polycrystalline nanofilms exceed the range of our findings: between the Sommerfeld value and the bulk Lorentz ratio. To fully understand these effects, further experiments need to be carefully performed. Moreover, the effect of the grain boundary on the electron-phonon interaction should be further investigated.

**Acknowledgments**

This work was supported by the National Natural Science Foundation of China (Grant No. 51676121). We would also like to thank Dr. Lan Dong from Shanghai Jiao Tong University for valuable discussion. Simulations were performed on the π 2.0 cluster supported by the Center for High Performance Computing at Shanghai Jiao Tong University

**Reference**

[1] E. C. Walter, K. Ng, M. P. Zach, R. M. Penner, and F. Favier, Microelectron. Eng. **61**, 555 (2002).

[2] Y. Xiong and X. Lu, *Metallic Nanostructures* (Springer, Switzerland, 2015).

[3] A. Facchetti and T. J. Marks, *Transparent Electronics: From Synthesis to Applications* (John Wiley & Sons, 2010).

[4] A. L. Moore and L. Shi, Mater. Today **17**, 163 (2014).

[5] S. J. Mason, D. J. Wesenberg, A. Hojem, M. Manno, C. Leighton, and B. L. Zink, Phys. Rev. Mater. **4**, 065003 (2020).

[6] L. Qiu, N. Zhu, Y. Feng, E. E. Michaelides, G. Żyła, D. Jing, X. Zhang, P. M. Norris, C. N. Markides, and O. Mahian, Phys. Rep. **843**, 1 (2020).

[7] H. Wang, J. Liu, X. Zhang, and K. Takahashi, Int. J. Heat Mass Transf. **66**, 585 (2013).

[8] W. Ma and X. Zhang, Int. J. Heat Mass Transf. **58**, 639 (2013).

[9] S. D. Sawtelle and M. A. Reed, Phys. Rev. B **99**, 054304 (2019).

[10] D. Kojda, R. Mitdank, M. Handwerg, A. Mogilatenko, M. Albrecht, Z. Wang, J. Ruhhammer, M. Kroener, P. Woias, and S. F. Fischer, Phys. Rev. B **91**, 024302 (2015).

[11] Y. Zhao, M. L. Fitzgerald, Y. Tao, Z. Pan, G. Sauti, D. Xu, Y. Xu, and D. Li, Nano Lett. **20**, 7389 (2020).

[12] T. Q. Qiu and C. L. Tien, J. Heat Transfer **115**, 842 (1993).

[13] B. Feng, Z. Li, and X. Zhang, Thin Solid Films **517**, 2803 (2009).

[14] B. Feng, Z. Li, and X. Zhang, J. Phys. D. Appl. Phys. **42**, 055311 (2009).

[15] H. Lin, S. Xu, C. Li, H. Dong, and X. Wang, Nanoscale **5**, 4652 (2013).

[16] Z. Cheng, L. Liu, S. Xu, M. Lu, and X. Wang, Sci. Rep. **5**, 10718 (2015).

[17] M. Yarifard, J. Davoodi, and H. Rafii-Tabar, Comput. Mater. Sci. **111**, 247 (2016).

[18] A. D. Avery, S. J. Mason, D. Bassett, D. Wesenberg, and B. L. Zink, Phys. Rev. B **92**,

214410 (2015).

[19] G. Chen, *Nanoscale Energy Transport and Conversion: A Parallel Treatment of Electrons, Molecules, Phonons, and Photons* (Oxford University Press, 2005).

[20] D. G. Cahill, P. V. Braun, G. Chen, D. R. Clarke, S. Fan, K. E. Goodson, P. Keblinski, W. P. King, G. D. Mahan, A. Majumdar, H. J. Maris, S. R. Phillpot, E. Pop, and L. Shi, Appl. Phys. Rev. **1**, 011305 (2014).

[21] J. J. Plombon, E. Andideh, V. M. Dubin, and J. Maiz, Appl. Phys. Lett. **89**, 113124 (2006).

[22] T. Sun, B. Yao, A. P. Warren, K. Barmak, M. F. Toney, R. E. Peale, and K. R. Coffey, Phys. Rev. B **79**, 041402 (2009).

[23] C. Shao, Q. Rong, N. Li, and H. Bao, Phys. Rev. B **98**, 155418 (2018).

[24] K. Fuchs, Math. Proc. Cambridge Philos. Soc. **34**, 100 (1938).

[25] A. F. Mayadas and M. Shatzkes, Phys. Rev. B **1**, 1382 (1970).

[26] X. Zhang, H. Xie, M. Fujii, H. Ago, K. Takahashi, T. Ikuta, H. Abe, and T. Shimizu, Appl. Phys. Lett. **86**, 171912 (2005).

[27] Q. G. Zhang, B. Y. Cao, X. Zhang, M. Fujii, and K. Takahashi, Phys. Rev. B **74**, 134109 (2006).

[28] N. Stojanovic, D. H. S. Maithripala, J. M. Berg, and M. Holtz, Phys. Rev. B **82**, 075418 (2010).

[29] Q. G. Zhang, B. Y. Cao, X. Zhang, M. Fujii, and K. Takahashi, Phys. Rev. B **74**, 134109 (2006).

[30] G. V. Chester and A. Thellung, Proc. Phys. Soc. **77**, 1005 (1961).

[31] F. Völklein, H. Reith, T. W. Cornelius, M. Rauber, and R. Neumann, Nanotechnology **20**, 325706 (2009).

[32] F. Völklein, H. Reith, T. W. Cornelius, M. Rauber, and R. Neumann, Nanotechnology **20**, 325706 (2009).

[33] D. Choi, C. S. Kim, D. Naveh, S. Chung, A. P. Warren, N. T. Nuhfer, M. F. Toney, K. R. Coffey, and K. Barmak, Phys. Rev. B **86**, 045432 (2012).

[34] D. Gall, J. Appl. Phys. **127**, 050901 (2020).

[35] Y. Chen, J. Ma, and W. Li, Phys. Rev. B **99**, 020305 (2019).

[36] Z. Tong, S. Li, X. Ruan, and H. Bao, Phys. Rev. B **100**, 144306 (2019).

[37] S. Li, Z. Tong, X. Zhang, and H. Bao, arXiv Preprint arXiv:2004.08843, 2020.

[38] S. Li, A. Wang, Y. Hu, X. Gu, Z. Tong, and H. Bao, Mater. Today Phys. **15**, 100256 (2020).

[39] B. Fu, K. D. Parrish, H. Y. Kim, G. Tang, and A. J. H. McGaughey, Phys. Rev. B **101**, 045417 (2020).

[40] S. Bose, C. Galande, S. P. Chockalingam, R. Banerjee, P. Raychaudhuri, and P. Ayyub, J. Phys. Condens. Matter **21**, 205702 (2009).

[41] R. C. Munoz and C. Arenas, Appl. Phys. Rev. **4**, 011102 (2017).

[42] J. Y. Murthy and S. R. Mathur, J. Heat Transfer **124**, 1176 (2002).

[43] Y. Hu, T. Feng, X. Gu, Z. Fan, X. Wang, M. Lundstrom, S. S. Shrestha, and H. Bao, Phys. Rev. B **101**, 155308 (2020).

[44] D. Josell, S. H. Brongersma, and Z. Tőkei, Annu. Rev. Mater. Res. **39**, 231 (2009).

[45] S. B. Soffer, J. Appl. Phys. **38**, 1710 (1967).

[46] J. R. Sambles, Thin Solid Films **106**, 321 (1983).

[47] R. L. Graham, G. B. Alers, T. Mountsier, N. Shamma, S. Dhuey, S. Cabrini, R. H. Geiss, D. T. Read, and S. Peddeti, Appl. Phys. Lett. **96**, 23 (2010).

[48] T. Sun, B. Yao, A. P. Warren, K. Barmak, M. F. Toney, R. E. Peale, and K. R. Coffey, Phys. Rev. B **81**, 155454 (2010).

[49] H. K. Liu, Y. Lin, and S. N. Luo, J. Phys. Chem. C **118**, 24797 (2014).

[50] C. Kimmer, S. Aubry, A. Skye, and P. K. Schelling, Phys. Rev. B **75**, 144105 (2007).

[51] A. Bietsch and B. Michel, Appl. Phys. Lett. **80**, 3346 (2002).

[52] J. S. Chawla, F. Gstrein, K. P. O'Brien, J. S. Clarke, and D. Gall, Phys. Rev. B **84**, 235423 (2011).

[53] Z. Wang, J. E. Alaniz, W. Jang, J. E. Garay, and C. Dames, Nano Lett. **11**, 2206 (2011).

[54] C. Shao and H. Bao, Int. J. Heat Mass Transf. **85**, 33 (2015).

[55] A. Lavasani, D. Bulmash, and S. Das Sarma, Phys. Rev. B **99**, 085104 (2019).

[56] R. B. Ganvir, P. V. Walke, and V. M. Kriplani, Renew. Sustain. Energy Rev. **75**, 451 (2017).

[57] T. M. Tritt, *Thermal Conductivity: Theory, Properties, and Applications* (Springer

Science & Business Media, 2005).

[58] D. A. Broido, M. Malorny, G. Birner, N. Mingo, and D. A. Stewart, Appl. Phys. Lett. **91**, 231922 (2007).

[59] See supplementary material for details on (i) the deviation of the expression of thermal conductivity, (ii) average mean free path of Au and W and (iii) results for nanowires.

[60] J.M. Ziman, *Electrons and Phonons: The Theory of Transport Phenomena in Solids* (Oxford university press, 2001).

[61] Y. Wang, Z. Lu, and X. Ruan, J. Appl. Phys. **119**, 225109 (2016).

[62] A. K. Vallabhaneni, D. Singh, H. Bao, J. Murthy, and X. Ruan, Phys. Rev. B **93**, 125432 (2016).

[63] G. D. Mahan, *Many-Particle Physics* (Springer Science & Business Media, 2000).

[64] P. Giannozzi, S. Baroni, N. Bonini, M. Calandra, R. Car, C. Cavazzoni, D. Ceresoli, G. L. Chiarotti, M. Cococcioni, I. Dabo, A. Dal Corso, S. De Gironcoli, S. Fabris, G. Fratesi, R. Gebauer, U. Gerstmann, C. Gougoussis, A. Kokalj, M. Lazzeri, L. Martin-Samos, N. Marzari, F. Mauri, R. Mazzarello, S. Paolini, A. Pasquarello, L. Paulatto, C. Sbraccia, S. Scandolo, G. Sclauzero, A. P. Seitsonen, A. Smogunov, P. Umari, and R. M. Wentzcovitch, J. Phys. Condens. Matter **21**, 395502 (2009).

[65] J. P. Perdew and Y. Wang, Phys. Rev. B **98**, 195128 (2018).

[66] J. P. Perdew, K. Burke, and M. Ernzerhof, Phys. Rev. Lett. **78**, 1396 (1997).

[67] W. Li, J. Carrete, N. A. Katcho, and N. Mingo, Comput. Phys. Commun. **185**, 1747 (2014).

[68] J. Noffsinger, F. Giustino, B. D. Malone, C. H. Park, S. G. Louie, and M. L. Cohen, Comput. Phys. Commun. **181**, 2140 (2010).

[69] Y. S. Touloukian, *Thermophysical properties of matter* **1** (1970).

[70] F. Yang and C. Dames, Phys. Rev. B **87**, 035437 (2013).

[71] S. Zahiri, Z. Xu, Y. Hu, H. Bao, and Y. Shen, Int. J. Heat Mass Transf. **138**, 267 (2019).

[72] P. Zheng and D. Gall, J. Appl. Phys. **122**, 135301 (2017).

[73] E. Fawcett and D. Griffiths, J. Phys. Chem. Solids **23**, 1631 (1962).

[74] C. Kittel and P. McEuen, *Introduction to Solid State Physics* (New York: Wiley, 1967).

[75] F. Bloch, Zeitschrift Für Phys. **52**, 555 (1929).

[76] E. Grüneisen, Ann. Phys. **408**, 530 (1933).

[77] R. Sultan, A. D. Avery, J. M. Underwood, S. J. Mason, D. Bassett, and B. L. Zink, Phys. Rev. B **87**, 214305 (2013).

# Supplementary Material

## for

## First-principles based analysis of thermal transport in metallic nanostructures: size effect and Wiedemann-Franz law

Yue Hu[1], Shouhang Li[1], Hua Bao[1,*]

[1]University of Michigan–Shanghai Jiao Tong University Joint Institute, Shanghai Jiao Tong University, Shanghai 200240, P. R. China

## S1. Derivation: Expression of thermal conductivity of thin films

Steady-state Boltzmann transport equation with the relaxation time approximation is expressed as:

$$\mathbf{v}\cdot\nabla f + \frac{\mathbf{F}}{m}\cdot\nabla_{\mathbf{v}} f = -\frac{f-f^0}{\tau},$$

where the $f$ is the distribution function; the $\mathbf{v}$ is the velocity; the $\mathbf{F}$ is the external force; the $\tau$ is the relaxation time. We introduce a deviation function $g = f - f^0$, then the Boltzmann transport equation can be written as:

$$\mathbf{v}\cdot\nabla g + \mathbf{v}\cdot\nabla f^0 + \frac{\mathbf{F}}{m}\cdot\nabla_{\mathbf{v}} g + \frac{\mathbf{F}}{m}\cdot\nabla_{\mathbf{v}} f^0 = -\frac{g}{\tau}. \quad \text{(S1)}$$

In the above equation, we can assume that $\nabla_{\mathbf{v}} g = 0$ on the basis that the spatial size effect

---

[*]Author to whom correspondence should be addressed. *E-mail address*: hua.bao@sjtu.edu.cn (H. Bao).

does not affect $f$ in moment space [1]. The equation can be further simplified:

$$\tau \mathbf{v}\cdot\nabla g + g = -\frac{\mathbf{F}}{m}\cdot\nabla_v f^0 - \mathbf{v}\cdot\nabla f^0 = -S_0$$

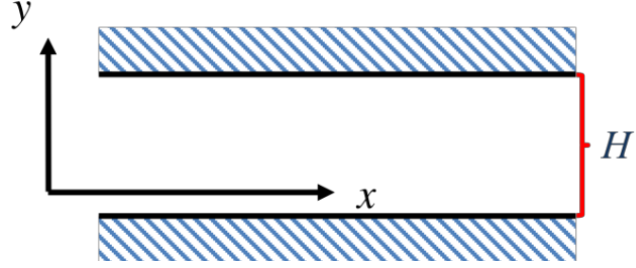

Fig. S1. Coordinates in thin film

For the thin film in Fig. S1, the solution of Eq. (S1) is [1]

$$g^+ = -S_0\left(1+G^+\exp\left(-\frac{y}{\tau|v_y|}\right)\right), v_y > 0$$

$$g^- = -S_0\left(1+G^-\exp\left(-\frac{d-y}{\tau|v_y|}\right)\right), v_y < 0$$

By applying the boundary condition ($p$ is sepcularity):

$$\begin{aligned} g^- &= pg^+, y=d \\ g^+ &= pg^-, y=0 \end{aligned},$$

we can obtain that $G^+ = G^- = \dfrac{1-p}{1-p\exp(-d/\tau|v_y|)}$. The heat flux is obtained by

$$q_x = \frac{1}{d}\int_y \sum_\lambda v_x g\hbar\omega dy.$$ When applying the temperature gradient along the thin film,

$S_0 = \dfrac{\partial T}{\partial x}\dfrac{\partial f^0}{\partial T}\tau v_x$. Thus the heat flux is:

$$q_x = \frac{\partial T}{\partial x}\sum_\lambda \hbar\omega\frac{\partial f^0}{\partial T}\tau'_\lambda v^2_{x,\lambda}S_\lambda, S_\lambda = \frac{1}{H}\int_0^H\left(1-\frac{1-p_\lambda}{1-p_\lambda\exp(-L/\tau'_\lambda|v_{y,\lambda}|)}\exp\left(-\frac{|y-y_B|}{\tau'_\lambda|v_{y,\lambda}|}\right)\right)dy.$$

According to the definition of the thermal conductivity: $k = \dfrac{q_x}{\partial T/\partial x}$, the expression of thermal conductivity is $\kappa = \sum_\lambda C_\lambda \tau'_\lambda v^2_{x,\lambda} S_\lambda$. According to Mayadas' derivation, when grain boundary exists, the $\tau'_\lambda$ is obtained by $\dfrac{1}{\tau'_\lambda} = \dfrac{1}{\tau_\lambda} + \dfrac{\alpha}{2\tau}\dfrac{v_\lambda}{|v_{\lambda,x}|}, \alpha = \dfrac{v_\lambda \tau_\lambda}{D}\dfrac{R}{1-R}$. When the carrier is free electron,

our expression can be equivalence to the expression in Mayadas' publication [2].

## S2. Expression of thermal conductivity of nanowires

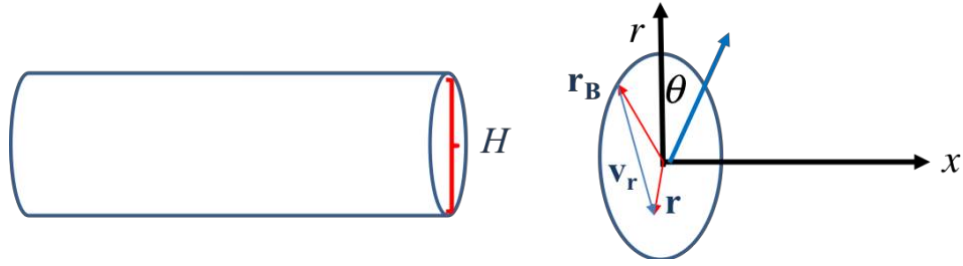

Fig. S2. Coordinates in Nanowires

For the nanowire in Fig. S2, the solution of Eq. (S1) is [3]:

$$g = -S_0\left(1-\exp\left(-\frac{|\mathbf{r}-\mathbf{r}_B|}{\tau|v_r|}\right)\right).$$

Similar to the deviation, the expression of thermal conductivity in nanowires is:

$$\kappa = \sum_\lambda C_\lambda \tau'_\lambda v_{x,\lambda}^2 S_\lambda, \quad \frac{1}{\tau'_\lambda} = \frac{1}{\tau_\lambda} + \frac{\alpha}{2\tau}\frac{v_\lambda}{|v_{\lambda,x}|}, \quad \alpha = \frac{v_\lambda \tau_\lambda}{D}\frac{R}{1-R}.$$

$$S_\lambda = \frac{4}{\pi H^2}\iint_0^D\int_{2\pi}\left(1-\frac{1-p_\lambda}{1-p_\lambda \exp\left|-\frac{2\sqrt{\left(\frac{D}{2}\right)^2-r^2\sin^2(\phi-\theta)}}{\tau'_\lambda \|\mathbf{v}_{\lambda,r}\|}\right|}\times \exp\left(-\frac{r\cos(\phi-\theta)+\sqrt{\left(\frac{D}{2}\right)^2-r^2\sin^2(\phi-\theta)}}{\tau_i\|\mathbf{v}_{\lambda,r}\|}\right)\right)rd\theta dr$$

## S3. Average mean free path of Au and W from 100K to 400K

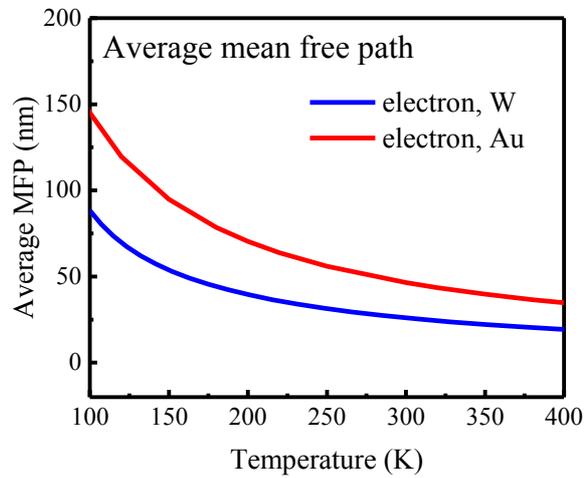

Fig. S3. (a) Averaged electron mean free path for Au and W

## S4. Size effect in nanowires

### a) Electron contribution

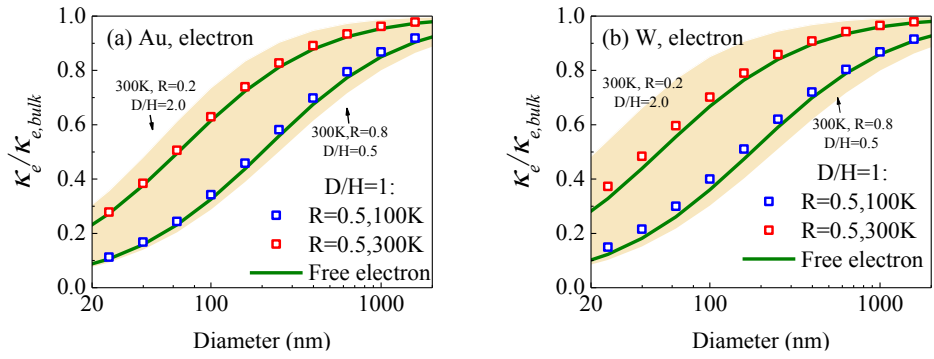

Fig. S4. Diameter $H$ dependent electronic thermal conductivity over bulk electronic thermal conductivity for (a) Au nanowire and (b) W nanowire.

### b) Phonon contribution

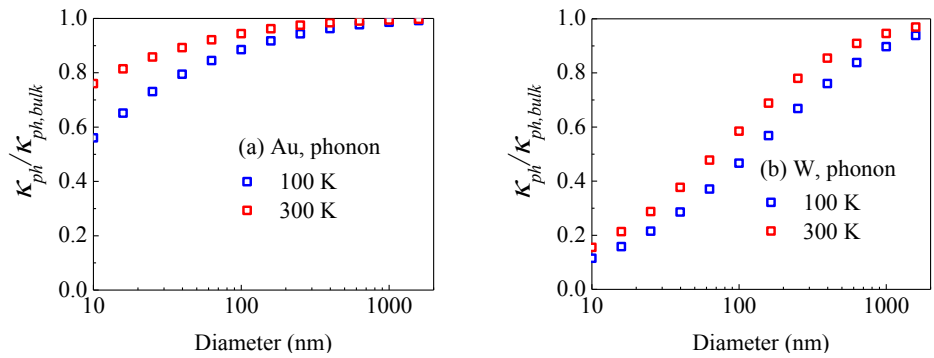

Fig. S5. Diameter $H$ dependent phonon thermal conductivity over bulk phonon thermal conductivity for (a) Au nanowire and (b) W nanowire.

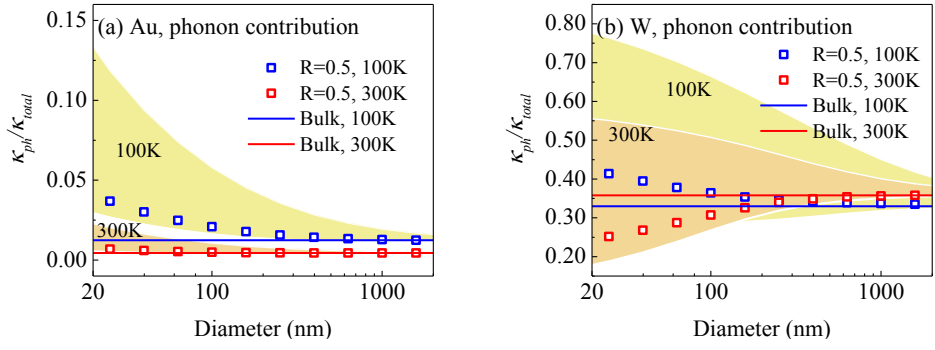

Fig. S6. Diameter $H$ dependent percentage of phonon thermal conductivity contributing to the total thermal conductivity for (a) Au nanowire and (b) W nanowire.

## S5. Lorentz ratio in nanowires

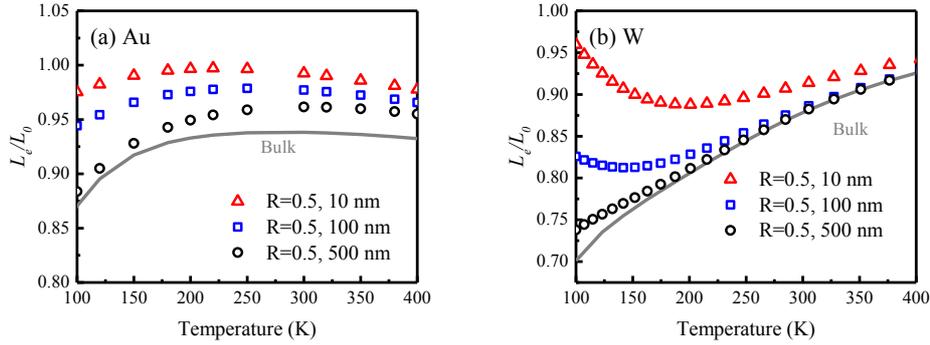

Fig. S7. Temperature dependent Lorentz ratio $L_e$ for (a) Au nanowire and (b) W nanowire. The Lorentz ratio is normalized by Sommerfeld value, as $L_0 = 2.44 \times 10^{-8} \, \text{W}\Omega/\text{K}^2$.

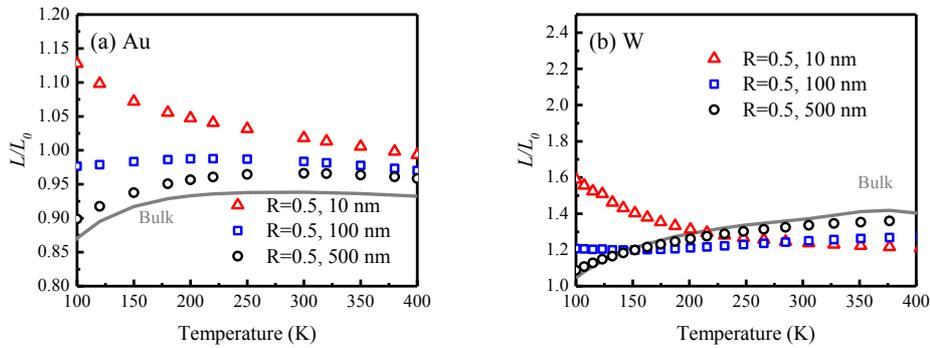

Fig. S8. Temperature dependent Lorentz ratio $L$ for (a) Au nanowire and (b) W nanowire. The

Lorentz ratio is normalized by Sommerfeld value, as $L_0 = 2.44 \times 10^{-8} \, \text{W}\Omega/\text{K}^2$.